\documentclass[sigconf,natbib=true]{acmart} %
\usepackage{graphicx} %
\usepackage[show]{chato-notes}
\usepackage{amsmath}
\usepackage{algorithm}
\usepackage{subfig}
\usepackage{algpseudocode}
\usepackage{url}
\usepackage{makecell}
\usepackage{subfig}
\usepackage{enumitem}
\usepackage{placeins}
\usepackage{multirow}
\usepackage{soul}

\usepackage[para]{footmisc}

\usepackage{marginnote}
\newcommand{\pageenlarge}[1]{\enlargethispage{#1\baselineskip}}

\newcommand\m[1]{\begin{bmatrix}#1\end{bmatrix}} 
\newcommand{\jpqprune}{\textsf{RecJPQPrune}}

\newcommand{\sasha}[1]{\textcolor[HTML]{000000}{#1}}

\newcommand{\craig}[1]{\textcolor{black}{#1}}
\newcommand{\nt}[1]{\textcolor{black}{#1}}
\newcommand{\sg}[1]{\textcolor{black}{#1}}
\newcommand{\sgw}[1]{\textcolor[HTML]{000000}{#1}}
\newcommand{\sgt}[1]{\textcolor[HTML]{000000}{#1}}
\newcommand{\cm}[1]{\textcolor{black}{#1}}
\newcommand{\sgf}[1]{\textcolor[HTML]{000000}{#1}}
\newcommand{\sgff}[1]{\textcolor[HTML]{000000}{#1}}
\newcommand{\crc}[1]{\textcolor[HTML]{000000}{#1}}

\usepackage{fontawesome5}

\newcommand{\argmax}{\operatornamewithlimits{\sf argmax}}

\usepackage{xcolor}
\usepackage{tcolorbox}

\makeatletter
\newcommand*{\transpose}{\bgroup\@ifstar{\mathpalette\@transpose{\mkern-3.5mu}\egroup}{\mathpalette\@transpose\relax\egroup}}
\newcommand*{\@transpose}[2]{\setbox0=\hbox{\m@th$#1#2\intercal$}\raise\dp0\box0}
\makeatother

\title[Efficient Recommendation with Millions of Items by Dynamic Pruning of Sub-Item Embeddings]{Efficient Recommendation with Millions of Items by \\ Dynamic Pruning of Sub-Item Embeddings}

\copyrightyear{2025}
\acmYear{2025}
\setcopyright{cc}
\setcctype{by}
\acmConference[SIGIR '25]{Proceedings of the 48th International ACM SIGIR Conference on Research and Development in Information Retrieval}{July 13--18, 2025}{Padua, Italy}
\acmBooktitle{Proceedings of the 48th International ACM SIGIR Conference on Research and Development in Information Retrieval (SIGIR '25), July 13--18, 2025, Padua, Italy}\acmDOI{10.1145/3726302.3729963}
\acmISBN{979-8-4007-1592-1/2025/07}

\ccsdesc[500]{Information systems~Information retrieval}

\keywords{Dynamic Pruning, Recommendation Systems}

\author{Aleksandr V. Petrov}
\affiliation{%
  \institution{University of Glasgow} \country{United Kingdom}}

\email{a.petrov.1@research.gla.ac.uk}

\author{Craig Macdonald}
\affiliation{%
  \institution{University of Glasgow} \country{United Kingdom}}
\email{craig.macdonald@glasgow.ac.uk}

\author{Nicola Tonellotto}
\affiliation{%
  \institution{University of Pisa} \country{Italy}}
\email{nicola.tonellotto@unipi.it}

\begin{document}

\begin{abstract}

\looseness -1 A large item catalogue is \crc{a major challenge}
for deploying modern sequential recommender models, \crc{since} it makes the memory footprint of the model large and increases inference latency. One promising approach to address this is RecJPQ, which replaces item embeddings with sub-item embeddings. However, slow inference remains problematic because finding the top highest-scored items usually requires scoring all items in the catalogue, which may not be feasible for large catalogues. By adapting dynamic pruning concepts from \crc{document} retrieval, we propose the \jpqprune{} dynamic pruning algorithm to efficiently find the top highest-scored items without computing the scores of all items in the catalogue. Our \jpqprune\ algorithm is \emph{safe-up-to-rank $K$} since it theoretically guarantees that no potentially high-scored item is excluded from the final top $K$ recommendation list, \crc{thereby ensuring no impact on effectiveness}. \cm{Our experiments on two large datasets and three recommendation models demonstrate the efficiency achievable using \jpqprune{}: for instance, on the Tmall dataset with 2.2M items, we can reduce the median model scoring time by 64$\times$ compared to the Transformer Default baseline, and 5.3$\times$ compared to a recent scoring approach called PQTopK. Overall, this paper demonstrates the effective and efficient inference of Transformer-based recommendation models at catalogue scales not previously reported in the literature. \crc{Indeed, our \jpqprune{}\ algorithm can score 2 million items in under 10 milliseconds without GPUs, and without relying on Approximate Nearest Neighbour (ANN) techniques.}}\pageenlarge{1}

\end{abstract}

\maketitle

\section{Introduction} \label{sec:intro}\pageenlarge{1}
\looseness -1 Sequential recommender systems use lists of user-item interactions to predict the next item \sgf{a user is likely to interact with}. Using the ordering of interactions allows a recommender system to account for evolving user interests as well as sequential patterns in item interactions common in real-world applications, e.g., users prefer to watch a movie series from the first to the last.
The \emph{next item prediction} task tackled by sequential recommender systems is similar to the \next{next token prediction} task addressed by language models. Therefore, models that were initially developed for language modelling have been successfully adapted for sequential recommendation. In particular, Transformer-based models~\cite{Transformer}, \sgf{such as models based on} BERT4Rec~\cite{BERT4Rec, CBiT} and SASRec~\cite{SASRec, petrovGSASRecReducingOverconfidence2023} \sgf{have achieved} state-of-the-art results for the next item prediction task. To solve the next item prediction task, these models use item ids in place of token ids. 

\looseness -1 However, despite similarities with language modelling, the number of items in the catalogue of a typical industrial recommender system may be several orders of magnitude larger compared with the vocabulary size of a typical language model~\cite{10.1145/3626772.3657816, zhaiActionsSpeakLouder2024}. \sasha{In particular,} Transformer-based language models, such as BERT~\cite{bert}, typically have \sasha{vocabularies} of less than \nt{50},000 tokens. In contrast, the number of items in the catalogues of recommender systems can reach hundreds of millions of items on large-scale platforms, such as Amazon or YouTube~\cite{youtube_videos, AmazonStatisticsUptoDate}. 
A large catalogue results in problems such as large GPU memory requirements for storing item embeddings, slow training and slow model inference~\cite{petrovRecJPQTrainingLargeCatalogue2024, Bert4RecRepro}.

\looseness -1 This paper specifically focuses on the \emph{slow inference problem}: a typical recommender system works using the "score-and-rank" approach, where the model first scores \emph{all} items in the catalogue and then selects the top $K$ items according to their scores.
 In contrast, in this paper, we consider a scenario in which scoring all items in the catalogue is impractical due to the large catalogue size. \sg{Existing methods for \sgf{avoiding exhaustive scoring} rely on heuristics, such as Approximate Nearest Neighbours (ANN)~\cite{chenApproximateNearestNeighbor2022}. However, these heuristics are \emph{unsafe}, meaning that they do not provide theoretical guarantees that all highly-scored items are included in the recommendation list, \cm{and hence can result in degraded effectiveness}. Moreover, most ANN-based methods require \crc{full} item embeddings to be trained in the first place, which may not be feasible in the large catalogue scenario~\cite{petrovRecJPQTrainingLargeCatalogue2024}.} Therefore, our goal is to develop a \sg{\emph{safe}} method that guarantees no degradation in recommendation effectiveness by finding the \sg{exact} top $K$ highest-scoring items \sg{while} only scoring \emph{some} \sgf{items in the catalogue}.  To achieve this, we build upon salient characteristics of \sg{RecJPQ~\cite{petrovRecJPQTrainingLargeCatalogue2024} -- a recently proposed embedding compression method for recommender systems.}

\sg{RecJPQ splits item ids into} a limited number of shared \emph{sub-item ids}, akin to how language models split words into shared subword tokens. Item embeddings can then be constructed as a concatenation of the sub-item embeddings, which results in substantially smaller models compared to storing full item embeddings. While helping to compress the model, the method does not \sg{fully} solve the slow inference problem. \sg{Indeed, even the PQTopK~\cite{petrovEfficientInferenceSubItem2024} algorithm (the most optimised version of the RecJPQ-based inference algorithm to date)} still applies the score-and-rank approach, which requires scoring all items.
However, we observe that the way \sg{PQTopK} computes item scores given \nt{the} scores of its sub-items is similar to how traditional information retrieval (IR) models compute document scores given a query: \nt{an} item score can be computed as the sum of the scores of the individual sub-items, which is similar to computing document score as a sum of token scores in the "bag-of-word" retrieval method, such as BM25~\cite{robertsonOkapiTRECRobertson1994}. 

\pageenlarge{2}\looseness -1 This inspires us to examine the applicability of {\em dynamic pruning} techniques that are typically employed to increase the efficiency of bag-of-word retrieval methods. Indeed, in this work, we propose \nt{the} pruning-based \jpqprune{} method for efficient calculation of top $K$ ranked items under \sg{RecJPQ}. The idea of this \crc{novel} method is based on the hypothesis that highly-ranked items should also have highly-scored sub-items. Instead of {\em exhaustively} computing scores for all items in the catalogue, we only compute scores for items that are associated with the highest-scored sub-items. Moreover, the specifics of sub-item representations in \sg{RecJPQ} allow us to compute an upper bound for the item scores, which allows us to stop the item scoring way before all items in the catalogue have been scored.

\looseness -1 \sgff{Recent generative recommender systems, such as TIGER~\cite{rajputRecommenderSystemsGenerative2023} and GPTRec~\cite{GPTRec}, also rely on sub-item representations and avoid exhaustive item scoring by generating item ids autoregressively. However, both TIGER and GPTRec mention generation speed as a limitation, as they require a Transformer invocation for every generated sub-id, which makes them inefficient for retrieval and outside the scope of this paper.}

\looseness -1 \cm{In summary, this work contributes: (1) a novel dynamic pruning approach, \jpqprune{}, for speeding up inference for large-catalogue RecJPQ-based recommender models, with no impact on effectiveness; (2) experiments examining median and tail scoring times on two large datasets with millions of items, and for three Transformer-based sequential recommender models; and (3) a study into factors affecting the efficiency of \jpqprune{} for different models and different users. \jpqprune{} provides marked efficiency benefits -- 
for instance, on the Tmall dataset with 2.2M items, we can reduce the median model scoring time by 64$\times$ compared to default Transformer scoring, and 5.3$\times$ compared to PQTopK.}

\looseness -1 The structure of this paper is as follows: Section~\ref{sec:pruning} outlines existing applications of pruning in search and other machine learning scenarios. Section~\ref{sec:rec} describes \nt{state-of-the-art techniques for large-scale Transformer-based sequential recommendation models.} Section~\ref{sec:ours} describes \jpqprune{}. Our experimental setup, and our empirical validation demonstrating the benefits of \jpqprune{} follow in Sections~\ref{sec:exp} \&~\ref{sec:results}. \cm{To explain the variance between models and between users, in Section~\ref{sec:difficulty} we make a first examination of pruning difficulty in \jpqprune{}.} Section~\ref{sec:conc} provides concluding remarks. 

\section{\sasha{Pruning in Document Retrieval}}\label{sec:pruning}
\sasha{In the classical document retrieval task, the goal of the IR system is to retrieve textual \emph{documents} from a document collection $\mathcal{D}$ that are estimated most likely to be \emph{relevant} to a given textual query $q$. We focus on "bag-of-word" retrieval approaches, as exemplified by BM25~\cite{robertsonOkapiTRECRobertson1994}, in which both the documents $d \in \mathcal{D}$  and  the query $q$ are represented as multisets of \emph{terms} $t$.} 
Bag-of-word approaches compute query-document relevance estimates as:
\begin{align}
    \text{score}(q, d) = \sum_{t\in q} w(t,d) \label{eq:ir_doc_score}
\end{align}
\looseness -1 Each document has a non-negative integer known as a \textit{document identifier} (docid). Every term present in the collection has a \textit{posting list}, which comprises the docids of all documents where the term appears. The aggregated posting lists for all terms form the \textit{inverted index} of $\mathcal{D}$. %
The docids within a posting list can be arranged in ascending order, or by descending score/impact~\cite{anh01vector}.
The traditional approaches for processing queries and matching them to documents are the \textit{term-at-a-time} (TAAT) strategy, where the posting lists of query terms are processed sequentially, and the scores for each document are summed up in an \textit{accumulator} data structure; or \textit{document-at-a-time} (DAAT), where the posting lists of query terms are processed simultaneously while maintaining docid alignment.%

Processing queries exhaustively with TAAT or DAAT can be very inefficient. As a result, various dynamic pruning techniques~\cite{fntir2018} have been proposed,\footnote{\sgw{In the deep learning literature, the term \emph{pruning} typically means \emph{weight pruning}. In our case, we use the term pruning as used in IR literature~\cite{fntir2018} meaning pruning candidates from the scoring process to speed up scoring. For applications of weight pruning to recommender systems see, for example,~\cite{shenUMECUNIFIEDMODEL2021} or~\cite{dengDeepLightDeepLightweight2021}}.}, which \sgf{aim to} omit the scoring of (portions of) documents during query processing if they cannot make the final top $K$ retrieved set. Dynamic pruning strategies can be described as {\em safe-up-to-rank} $K$ -- meaning they are guaranteed to calculate the exact scores for each retrieved document, at least as deep as rank $K$ -- or {\em unsafe} -- indicating that their retrieval effectiveness may be negatively impact compared to an exhaustive scoring.

All dynamic pruning optimisations for TAAT involve a two-phase  strategy. In the initial phase, the TAAT algorithm is applied, processing one term at a time in ascending order of document frequency. %
New accumulators are created and updated until a pruning condition is satisfied. Subsequently, the second phase commences, during which no new accumulators are created~\cite{buckley:1985,Moffat94fastranking,moffat:zobel:1996}.

\pageenlarge{2} \looseness -1 Among the dynamic pruning strategies for DAAT, MaxScore~\cite{turtle:flood:1995a}, WAND~\cite{10.1145/956863.956944}, \nt{and their variants~\cite{ding:2011,vbmw}} stand out as the most widely used. Both approaches enhance the inverted index by recording the maximum score contribution for each term. This enables the safe skipping of substantial segments within posting lists if those segments only consist of terms whose combined maximum scores are lower than the scores of the top $K$ documents already identified during query processing, known as the threshold, and denoted $\theta$. They also utilise a global per-term upper bound, i.e., the maximum score across all documents containing the term, in order to make pruning decisions. 
\nt{Finally, there are a number of dynamic pruning strategies for impact-ordered posting lists, such as score-at-a-time~\cite{anh01vector}. Similarly, our approach  uses score-sorted ids to perform computations as
efficiently as possible~\cite{10.1145/3576922}, but we do not leverage related dynamic pruning techniques e.g.~anytime ranking~\cite{Zilberstein_1996}, due to their inherent unsafeness, leaving to future work the analysis of unsafe settings.}

\looseness -1 \nt{In this paper}, we propose a safe-up-to-rank-$K$ novel dynamic pruning strategy,  \jpqprune{}, which is a hybrid of both TAAT and DAAT dynamic pruning, but designed specifically for scoring RecJPQ item representations in recommender systems. We position \jpqprune{} within the dynamic pruning literature in Section~\ref{sec:ours}.

\looseness -1 Other works focussed on improving the efficiency of expensive machine-learned ranking models have addressed (i) early termination of regression trees~\cite{10.1145/1718487.1718538}, or, (ii) more recently, the early termination of layers in Transformer-based  models~\cite{xin-etal-2020-deebert}, such as cross-encoders. 
These optimisations are applied for each candidate document being ranked.
In our setting, the Transformer model needs to be applied only once to obtain a representation of the user's recommendation need, like a query encoder in neural dense document retrieval. Hence, approaches that speed up model inference are orthogonal \nt{to our proposed \jpqprune{} strategy}.

\section{Transformers for Large-catalogue Sequential Recommendation}\label{sec:rec}
\looseness -1 In this section, we cover the background of the large-scale sequential recommendation. Section~\ref{ssec:recsys:preliminarilies} provides an overview of current state-of-the-art models for \sg{large-scale} sequential recommendation. \sg{Section~\ref{sec:rejpq} details RecJPQ, \nt{the} state-of-the-art method for item embedding compression and PQTopK, an efficient scoring algorithm for RecJPQ-based models as a baseline for our work.}%

\subsection{Preliminaries} \label{ssec:recsys:preliminarilies}
\looseness -1 Sequential recommendation is seen as a \emph{next item prediction} task. Formally, given a chronologically ordered sequence of user-item interactions $h = \left\{i_1, i_2, i_3 ... i_n\right\}$, \nt{also known as their \textit{interactions history},} the goal of a recommender system is to predict the next item in the sequence, $i_{n+1}$ from the \nt{\emph{item catalogue} $\mathcal{I}$, that is, the set of all possible items.} The total number of items $|\mathcal{I}|$ is the \emph{catalogue size}.

\looseness -1 Sequential recommendation bears a resemblance to the natural language processing task of \emph{next token prediction}, as addressed by language models, such as GPT-2~\cite{gpt2}.
Hence, \cm{language models have been adapted to recommendation,} by replacing token ids (in language models) with item ids (in recommendation models). In particular, Transformer-based models~\cite{Transformer}, such as SASRec~\cite{SASRec} (based on the Transformer Decoder architecture, like GPT-2)  and BERT4Rec~\cite{BERT4Rec} (based on Transformer Encoder, like BERT) have \cm{exhibited} state-of-the-art results for sequential recommendation~\cite{Bert4RecRepro}.

\pageenlarge{2}Typically, to generate recommendations given a history of interactions 
$h$, 
Transformer-based models first generate a sequence embedding $\phi \in \mathbb{R}^d$, where $d$ is the embedding dimensionality, using the Transformer model, such that $\phi =  \textsf{Transformer}(h)$\footnote{In document retrieval parlance, this would be a query embedding.}.
The scores for all items, \nt{denoted as } $r = (r_1, \ldots, r_{|\mathcal{I}|}) \in \mathbb{R}^{|\mathcal{I}|}$, are then computed by multiplying the matrix of item embeddings $W \in \mathbb{R}^{|\mathcal{I}|\times d}$, \nt{where $d$ is the embeddings dimension,} shared with the embeddings layer of the Transformer model, by the sequence embedding $\phi$:\footnote{\cm{This generalises to models with item biases (e.g.\ BERT4Rec) if we assume that the first dimension of the sequence embedding is 1 and the first dimension of item embedding is the bias.}}
\begin{align}
r = W\phi \label{eq:scores} 
\end{align}

Finally, the model generates recommendations by selecting \sgf{from} $r$ the top $K$ items with the highest scores. \sg{In the rest of the paper, we refer to this simple matrix-to-vector multiplication-based scoring procedure as \emph{Transformer Default}. }%

\looseness -1 Despite their effectiveness, training Transformer-based models with large item catalogues is a challenging task, as these models typically have to be trained for long time~\cite{Bert4RecRepro} and require appropriate selection of training objective~\cite{petrovRSSEffectiveEfficient2023}, negative sampling strategy and loss function~\cite{petrovGSASRecReducingOverconfidence2023,klenitskiyTurningDrossGold2023}.
Transformer-based models with large catalogues also require a lot of memory to store the item embeddings $W$. This problem has recently been addressed by \sg{RecJPQ}, which we \sg{detail} in the next section. 
Finally, another problem of Transformer-based models with large catalogues is their slow inference with large catalogues. Indeed, computing all item scores using Equation~\eqref{eq:scores} may be prohibitively expensive when the item catalogue is large: it requires $|I|\times d$ scalar multiplications and additions, and, as we noted in Section~\ref{sec:intro},  in real-world recommender systems, the catalogue size $|I|$ may reach hundreds of millions of items, making exhaustive computation using Equation~\eqref{eq:scores} impractical. Moreover, typically large-catalogue recommender systems have a large number of users as well; therefore, the model has to be used for inference very frequently and, ideally, using only low-cost hardware, i.e., without GPU acceleration. Therefore, real-life recommender systems rarely exhaustively score all items for all users and instead apply unsafe heuristics (i.e.\ \cm{which} do not provide theoretical guarantees that all high-scored items will be returned), such as two-stage ranking. However, as we show in Section~\ref{sec:ours}, it is possible to return the top $K$ items exactly without scoring all items exhaustively for \sg{RecJPQ}-based recommendation models. We now discuss the \sg{RecJPQ approach itself and how it can be used for efficient model inference}.
 
\subsection{RecJPQ and PQTopK} \label{sec:rejpq}
\looseness -1 \pageenlarge{2} \cm{Product Quantisation (PQ)~\cite{jegouProductQuantizationNearest2011} is a family of methods for compressing embeddings tensors, where the full embeddings are decomposed into sub-embeddings. PQ \sgf{is an active area of research (e.g.~\cite{leeBRBKMeansEnhancing2024}) and it} is widely used in approximate nearest neighbour approaches, exemplified by the FAISS library~\cite{FAISS}. PQ requires that the full embeddings already exist; in contrast, RecJPQ~\cite{petrovRecJPQTrainingLargeCatalogue2024} is a solution for jointly training effective and compressed item embedding tensors in sequential recommender systems instead of training full embeddings and followed by embedding compression \sgf{(as done by e.g. EODRec~\cite{xiaEfficientOnDeviceSessionBased2023}, LightRec~\cite{lianLightRecMemorySearchEfficient2020} \& MDQE~\cite{wangCompressingEmbeddingTable2022})}. While RecJPQ is named after Joint Product Quantisation~\cite{zhanJointlyOptimizingQuery2021} (a dense retrieval technique), the main idea of RecJPQ is inspired by tokenisation in language models. Similar to how words are split into sub-word tokens in language models, item ids are split into sub-item ids in RecJPQ. RecJPQ can be used with many embedding-based sequential recommendation models (including state-of-the-art Transformer-based models, such as BERT4Rec). 
In this section, we provide an overview of RecJPQ and the RecJPQ-based PQTopK approach for efficient scoring~\cite{petrovEfficientInferenceSubItem2024}.}  

\looseness -1 For notational convenience, let $[N]$ denote the set $\{1, 2, \ldots, N\}$. Let $\mathcal{I}$ denote our item catalogue, where each item is identified with an item id $i \in [\mathcal{|I|}]$.
\sg{RecJPQ builds a \emph{codebook}} $\mathcal{G}$, \sg{which} is composed \sg{of} $M$ splits, and each split is composed \sg{of} $B$  distinct sub-item ids with associated embeddings in $\mathbb{R}^{d/M}$. 
The codebook plays two roles: first, given an item id, it returns a list of $M$ sub-item ids $g_{im}$, and second, given a split id and a sub-item id, it returns the corresponding embedding. Formally, these roles can be modelled as two functions:
\begin{align*}
    G_1: \; & \mathcal{I} \to [B]^M \; \text{such that} \; G_1(i) = (g_{i1}, g_{i2}, \ldots, g_{iM}),\\
    G_2: \; & [M] \times [B] \to \mathbb{R}^{d/M} \; \text{such that} \; G_2(m,g_{im}) =  \psi_{m,g_{im}}.
\end{align*}%
\looseness -1 The function $G_1$ can be implemented as a lookup table with $|\mathcal{I}|$ item ids, each pointing to $M$ sub-item ids. The function \nt{$G_2$} can be implemented as $M$ embedding matrices with $B$ rows of $d/M$ real values. \sg{RecJPQ constructs mapping $G_1$ using SVD decomposition of the user-item matrix and trains embeddings $G_2$ as part of \cm{the sequential recommendation model's normal training process}. \sgf{However, due to our focus on inference, the details of} model training are not important for this work, and we refer the reader seeking more details on the construction of these mappings to the original publication~\cite{petrovRecJPQTrainingLargeCatalogue2024}.}

Given an item id $i$, \sg{RecJPQ} \sgt{constructs} the corresponing item embedding $w_i$ by looking up the corresponding \sg{sub-item} ids $g_{i1}, \ldots, g_{iM}$ from $G_1$, and then, for each split $m$ and \sg{sub-item} id $g_{im}$, by using the $m$-th embedding matrix and its $g_{im}$-th row, i.e., 
\begin{align}
    w_i = \m{\psi_{1,g_{i1}}^\transpose|\psi_{2,g_{i2}}^\transpose|\cdots|\psi_{M,g_{iM}}^\transpose}^\transpose,   \label{eq:embedding_reconstruct} 
\end{align}
\sg{were "$|$" denotes concatenation, i.e., an item embedding in RecJPQ is obtained by concatenating the associated sub-item id embeddings.} \cm{After \sgt{obtaining} all item embeddings, the Transformer Default scoring procedure (Equation~\eqref{eq:scores}) can be applied.}

\looseness-1   \cm{Recently, a \sgt{refined} scoring procedure called PQTopK was proposed for RecJPQ-based models, offering superior efficiency to the Transformer Default approach~\cite{petrovEfficientInferenceSubItem2024}.}
In PQTopK, the input sequence embedding $\phi$ obtained from the Transformer is split into a list of $M$ sub-embeddings $\phi_1, \phi_2, \ldots, \phi_M$, with $\phi_{m} \in \mathbb{R}^{d/M}$ for $m \in [M]$, i.e.:
\begin{align*}
    \phi = \m{\phi_1^\transpose|\phi_2^\transpose|\cdots|\phi_M^\transpose}^\transpose,
\end{align*}
s.t.~the score $r_i$ for item $i$ \cm{is} the sum of sub-embedding dot-products: 
\begin{align}
   r_i = w_i \cdot \phi = \sum_{m =1}^M \psi_{m,g_{im}} \cdot \phi_m.
\end{align}
\pageenlarge{2}\sg{There \nt{are} $M$ splits, and $B$ embeddings in each split, meaning that there \nt{are}} $M\times B$ partial scores between $\psi_{m,b}$ and $\phi_m$. \sg{These partial scores can} be precomputed, and stored in a matrix $S \in \mathbb{R}^{M \times B}$, called \crc{the} \emph{sub-item id score matrix}.
Hence, $r_i$ can also be computed as:
\begin{align}\label{eq:sum_sub_scores}
   r_{i} = \sum_{m=1}^M S_{m,g_{im}}.
\end{align}
\nt{The precomputed matrix $S$ stores only $M Bd/M=Bd$ floats (in our experiments $Bd = 256  \cdot 512 = 131,072$), while the matrix $W$ stores $|\mathcal{I}|d \gg Bd$ floats (in our experiments $|\mathcal{I}|d \geq 512 \cdot 10^6$).} \sgw{Computation of this matrix is also very efficient; for example, in our experiments, the time required to compute this matrix was only 0.2ms, which is negligible compared to the time required to compute \crc{the} sequence embedding $\phi$ ($\sim$24ms for the SASRecJPQ backbone, see Table~\ref{tab:expmodels}). }
\sg{After computing the scores for every item, PQTopK ranks them \crc{by} score and returns the top $K$ items with the highest scores as recommendations.} \sg{As the sub-item id scores are computed once for each new user request and then reused when scoring all items in the catalogue, computing scores \nt{in} this way results in efficiency gains while returning the same list of items as Transformer Default.\footnote{\sgw{Note that the PQTopK algorithm can score not only full catalogue of the items but also an arbitrary subset of items; we use this property to score promising items in our proposed \jpqprune{}. PQTopK is also a fully vectorisable algorithm.  For more details of PQTopK and its pseudo-code, see~\cite{petrovEfficientInferenceSubItem2024}}.} Indeed, \sgt{the PQTopK paper}~\cite{petrovEfficientInferenceSubItem2024} reported 4.5$\times$ faster inference using PQTopK when compared to Transformer Default. \sgw{However, despite efficiency gains, retrieving the top $K$ items still requires computing the score of every item in the catalogue, limiting its applicability with very large catalogues where computing scores of every item may not be feasible.} In the next section, we propose an algorithm that addresses this limitation and finds the top $K$ items without scoring all items.}

\color{black}
\section{RecJPQPrune}\label{sec:ours}

\looseness -1 We now describe our \jpqprune{} method, discussing the principles we use for dynamic pruning of \sg{RecJPQ}-based \cm{representations} (Section~\ref{ssec:principles}), the algorithm itself (Section~\ref{ssec:alg}), \nt{and }its positioning \sgf{w.r.t.}\ existing IR dynamic pruning strategies (Section~\ref{ssec:related}). %

\subsection{\mbox{\hspace{-1mm}Dynamic Pruning Principles for RecJPQPrune}}\label{ssec:principles}
\looseness -1 \sasha{Our goal is to build an algorithm that allows us to find the top-ranked items while avoiding exhaustive catalogue scoring. In Section~\ref{sec:pruning}, we discussed that for document retrieval, a similar problem can be solved using pruning techniques.} Comparing how document scores are computed in bag-of-word \craig{document retrieval} models (Equation~\eqref{eq:ir_doc_score}) and how item scores are computed in PQTopK (Equation~\eqref{eq:sum_sub_scores}), we find parallels between the two: in both cases, the final entity score is computed a sum of individual sub-entity scores. However, addressing the efficient computation of item scores requires the construction of a novel dynamic pruning algorithm. \jpqprune{} is based on \sg{three principles} that allow the scoring of some items to be omitted. We now describe these principles, while, later in Section~\ref{ssec:related}, we compare and contrast  \jpqprune{} with existing dynamic pruning approaches. %

\begin{figure}[tb]
\centering
\caption{\looseness -1 Relation between item ranks and sub-item scores for user 82082 from Gowalla: SASRecJPQ model~\cite{petrovRecJPQTrainingLargeCatalogue2024} with 8 splits and 256 sub-item ids per split. We highlight items ranked at the top, middle and bottom of the ranking.
} \label{fig:gowalla_heatmap}
\resizebox{ \linewidth}{!}{
    \includegraphics{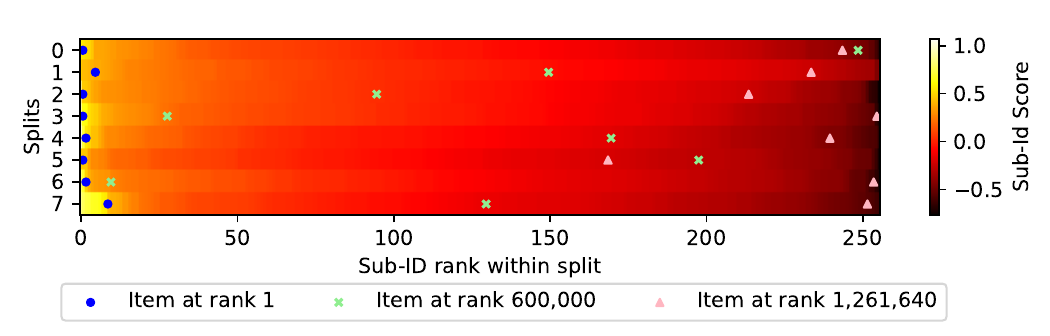}
}\vspace{-1.2em}
\end{figure}

\pageenlarge{2}\looseness -1 \paragraph{P1. Items with high scores typically have sub-items with high scores as well} The intuition behind this principle is that the total score of an item is calculated as the sum of its sub-item scores. Hence, a high overall score generally requires several high-scoring sub-items. Figure~\ref{fig:gowalla_heatmap} shows empirical evidence for this principle. The figure illustrates how sub-item id scores are distributed for user 82,082 in the Gowalla dataset computed by the SASRecJPQ model~\cite{petrovRecJPQTrainingLargeCatalogue2024}, \nt{where} the number of splits $M=8$ and the number of sub-items per split $B=256$. 
In the figure, the sub-item ids are ranked within their splits according to their score, with the highest-scored sub-item ids (bright-yellow colours) being on the left of the figure and the lowest-scored sub-item ids (dark-red \nt{colours}) being on the right of the figure. The figure highlights the sub-item ids associated with a top-ranked item, a middle-ranked item and a bottom-ranked item. As we can see, all sub-item ids of the top-ranked item appear to the left of the figure, i.e., they are scored high within their respective splits, and have bright colours, i.e., they are scored high across all sub-item ids. For the middle-scored item, we have a mixture of relatively high and relatively low-scored sub-item ids; however, none of the sub-item ids is the highest-scored sub-item id in the respective split. Most of the sub-item ids of the low-scored item also scored low. Overall, Figure~\ref{fig:gowalla_heatmap} supports Principle P1. 

In summary, Principle P1 suggests processing first the items associated with highly-scored sub-item ids during scoring before processing those linked to less highly-scored sub-item ids. This ensures we are likely to encounter all high-scored items relatively quickly. However, to achieve efficiency gains compared to exhaustive scoring, we need to be able to terminate scoring after all high-scored items have been found; to do that, we use Principle P2.

\pageenlarge{2} \looseness -1 \paragraph{P2. We can terminate scoring once the remaining \sasha{items} have no chance to enter the top $K$ results} \crc{Inspired by existing dynamic pruning techniques, we argue that, after processing a few sub-items, as described in Principle P1, it is possible to assert when any item in the remaining set of unscored items $\mathcal{I}^U \subset \mathcal{I}$ cannot enter the current top $K$ items.} \cm{Indeed,}  after a few iterations of scoring items associated with highly-scored sub-item ids, as described in Principle P1, we have, for each split $m$, a list $U_m$ of unprocessed sub-item ids; moreover, let $U = \{U_1, U_2, .. U_M\}$ denote all unprocessed sub-item ids across all $M$ splits. The \crc{unscored}, $\mathcal{I}^U$, are guaranteed to have all their sub-item ids appearing in $U$, because items associated with already processed sub-item ids were scored when their respective sub-item ids were processed. Therefore, from Equation~\eqref{eq:sum_sub_scores}, we can derive a score upper bound $\sigma$ for any unscored item $i \in I^U$:
\begin{align}\label{eq:upper_bound_strict}
    r_{i} &= \sum_{m=1}^M S_{m,i} \le \sum_{m=1}^M \max_{j \in U_m} S_{m,j} = \sigma
\end{align}
When processing sub-item ids as described in Principle P1, we can keep the minimum score within the current top $K$ highest scored items as a threshold~$\theta$. While the following \emph{pruning condition} holds 
\begin{align}\label{eq:pruning_condition}
    \sigma > \theta   
\end{align}
some items can still be entered into the top $K$ items. However, $\theta$ rises as items are admitted into the top $K$, and $\sigma$ falls as the unprocessed sub-item ids are less important. When the condition Equation~\eqref{eq:pruning_condition} no longer holds, we can guarantee that no item that has not been scored yet can enter into the top $K$ items; therefore, we can safely terminate the scoring algorithm. \sgw{In summary, Principle P2 argues that top $K$ items can be found without exhaustive scoring of all items in the catalogue, and provides us with the pruning condition that helps to identify the moment when scoring can be terminated.}%

\paragraph{P3.  Highly scored sub-item ids are frequently found in the same split.} In RecJPQ, sub-item ids are obtained using SVD decomposition of the user-item interaction matrix with different splits corresponding to different latent features of items. This means that if two items had similar values of a latent feature in the SVD decomposition, they would also have similar sub-item ids in the corresponding split. In short, in RecJPQ, \emph{similar items are assigned to similar sub-item ids}. As a result, in RecJPQ, highly scored sub-item ids are frequently clustered in the same split.
For example, looking again at Figure~\ref{fig:gowalla_heatmap}, we see that most of the highly-scored sub-item ids for the illustrated user are located in the \nt{last} split. \cm{This suggests} that once we find a promising sub-item id, we can process not only items associated with this sub-item id, but also items associated with other highly-scored sub-item ids from the same split. In other words, Principle P3  allows for \emph{batch processing} of sub-item ids.

\subsection{RecJPQPrune Algorithm}\label{ssec:alg}

Using Principles P1-P3, we can now derive the \jpqprune\ algorithm, \nt{illustrated in Algorithm~\ref{alg:jpqprune1}}. %

\looseness -1 According to Principle P1, \jpqprune\ processes sub-item ids in the descending order of their scores stored in $S$. \nt{The values in $S$ are computed efficiently at line~\ref{alg:jpqprune1:s}.}

\begingroup
\setlength{\textfloatsep}{0pt}
\begin{algorithm}[tb]
\small
\caption{\jpqprune{}($\phi$, $\mathcal{G}$, $\mathcal{L}$, $K$, $BS$)}\label{alg:jpqprune1}
\begin{algorithmic}[1]
    \Require Sequence embedding $\phi$
    \Require Codebook $\mathcal{G}$
    \Require Inverted indexes $\mathcal{L}_1, \ldots, \mathcal{L}_M$ %
    \Require Number of results to return $K$
    \Require Number of sub-item ids processed at every iteration $BS$ 

    \State $P \gets$ empty array of $M$ current sub-item positions, one per split
    \State $Q \gets$ empty array of $M$ empty sub-item id lists, one per split
    \State $S \gets$ compute the sub-item scores matrix \label{alg:jpqprune1:s} %

    \State $\sigma \gets 0$
    \color{teal}
    \newlength{\savedalgorithmicindent}
    \setlength{\savedalgorithmicindent}{\algorithmicindent}%
    \For{$m = 1, \ldots, M$} \label{alg:jpqprune1:init}
        \State $P[m] \gets 1$
        \State $Q[m] \gets$ sorted sub-item ids in split $m$ according to scores in $S$\label{alg:jpqprune1:sort}
        \State $\sigma \gets \sigma + S_{m,Q[m][1]}$
    \EndFor
    \color{black}
    \State $R_K \gets \text{empty list of $\langle$item id, score$\rangle$ pairs}$ 

    \State  $\theta \gets -\infty$  
    \While{$\sigma$ > $\theta$}  \label{alg:jpqprune1:mainloop}
        \State $m^* \gets \argmax_{ 1 \leq \text{m} \leq M} S_{m,Q[P[m]]}$\label{alg:jpqprune1:bestsplit}
        \State $I^* \gets \text{empty list of $\langle$item id, score$\rangle$ pairs}$
        \color{teal}
        \For{$j = 0, \ldots, BS-1$}\label{alg:jpqprune1:bs}
            \State $i^* \gets Q[m^*][P[m^*]+j]$
            \State $I^* \gets I^* \cup \mathcal{L}_{m^*}(i^*)$\label{alg:jpqprune1:retr}
        \EndFor
        \State $I_K \gets \texttt{PQTopK}(\mathcal{G}, S, K, I^*)$\label{alg:jpqrune1:score}
        \State $R_K \gets \texttt{merge}(R_K, I_K, K)$ \label{alg:jpqprune1:update_topk}
        \color{black}
        \State $P[m^*] \gets P[m^*] + BS$ \label{alg:jpqprune1:remove}

        \State $\sigma \gets 0$
        \color{teal}
        \For{$m = 1, \ldots, M$} \label{alg:jpqprune1:ubounds}
            \State $\sigma \gets \sigma + S_{m,Q[m][P[m]]}$\label{alg:jpqprune1:update_ub}
        \EndFor
        \color{black}
        \State $\theta \gets R_K[K].\text{score}$\label{alg:jpqprune1:update_th}
   \EndWhile
   \State\Return $R_K$ 
\end{algorithmic}
\end{algorithm}
\endgroup

\pageenlarge{2}\sgt{Then}, it sorts sub-item ids into the array $Q$ according to their scores within each split (line~\ref{alg:jpqprune1:sort}), \nt{using the array $P$ to track the position of the unprocessed sub-items in each split.}
\sgt{After that} it iterates \nt{through \sgt{the} splits in $Q$ and positions in $P$} \cm{while} the pruning condition holds (line~\ref{alg:jpqprune1:mainloop}).  At each iteration, it finds the maximum-scored unprocessed sub-item ids and corresponding split, denoted as $m^*$ (line~\ref{alg:jpqprune1:bestsplit}). From the split $m^*$, \jpqprune{} \cm{scores} $BS$ sub-item ids at a time, \nt{using the variable $i^*$} (line~\ref{alg:jpqprune1:bs}).
In order to be able to quickly score all items associated with the best split $m^*$ and the batch of best sub-item ids $i^*$, \jpqprune{} uses $M$ \emph{inverted indexes} $\mathcal{L}_1, \ldots, \mathcal{L}_M$. For a given split $m\in[M]$, the inverted index $\mathcal{L}_m$ maps a sub-item id to the set of all item id associated with the sub-item id (in effect, $\mathcal{L}_1, \ldots, \mathcal{L}_M$ are the inverse of $G_1$).
When processing $i^*$, \jpqprune{} retrieves all items associated with it from the inverted index $\mathcal{L}_{m^*}(i^*)$ (line~\ref{alg:jpqprune1:retr}). Then it computes \nt{the} scores of all items associated with this sub-item id using the PQTopK algorithm (line~\ref{alg:jpqrune1:score}), and updates \nt{the} current best top $K$ items (line~\ref{alg:jpqprune1:update_topk}).  Note that because every item is associated with multiple sub-item ids, \jpqprune{} may score some items multiple times; therefore, when updating current best top $K$ items \sgw{using the \texttt{merge} operation}, \jpqprune{} also deduplicates any repeated items. It then removes the \nt{sub-item ids in the batch} from the unprocessed sub-items (line~\ref{alg:jpqprune1:remove}), updates the upper bound $\sigma$ (line~\ref{alg:jpqprune1:update_ub}) and the pruning threshold $\theta$ (line~\ref{alg:jpqprune1:update_th}). \jpqprune{} iterates until the pruning condition (Equation~\eqref{eq:pruning_condition}) is met, after which it terminates and returns the current best top $K$ items.

\looseness -1 The use of batch processing addresses Principle P3, in that $BS$ \sg{sub-item ids} are identified at each outer loop \sgt{iteration}, and all items associated with these sub-item ids are processed in a single iteration. Following Principle P3, all these sub-item ids are taken from the same split $m^*$, enabling effective vectorisation of the for-loop at line~\ref{alg:jpqprune1:bs}. Moreover, the for-loops at lines~\ref{alg:jpqprune1:init} \&~\ref{alg:jpqprune1:ubounds}, \sgw{the PQTopK algorithm and the \texttt{merge} operation} are also vectorisable using common \cm{tensor} manipulation frameworks such as TensorFlow or PyTorch (all vectorisable operations are coloured \textcolor{teal}{teal} in Algorithm~1). Our TensorFlow implementation can be found in our \sgt{source code} repository.\footnote{\crc{Code for this paper: \faGithubSquare{}~\url{https://github.com/asash/recjpq_dp_pruning}}} 
\sg{\nt{The} batch size, $BS$, is an important parameter of the algorithm. On the one hand, larger batch sizes increase parallelism, hence making the algorithm more efficient. On the other hand, by increasing \nt{the} batch size, we score \nt{additional} items at every iteration, and hence, we may score more items than necessary before reaching the pruning condition. %
\crc{We empirically investigate batch size in Section~\ref{ssec:exp:bs}.}
}

\emph{Safety of \jpqprune{}.} \looseness -1 \jpqprune{} is a safe-up-to-rank $K$ dynamic pruning algorithm. Indeed, compared to the scoring of all items from all sub-items, the same exact scores are obtained to rank $K$ but minimising the processing of items that do not make the final top $K$. The safety is guaranteed by the fact that at termination time, the upper bound for scores for unprocessed items $\sigma$ is lower than the minimum score of the $K$ best items that the algorithm already found so that no unprocessed item can be included in the top $K$ items. \cm{Like existing dynamic pruning techniques, it is possible to make \jpqprune{} more efficient but unsafe (for instance by overinflating the threshold $\theta$ \crc{or limiting the number the iterations}), resulting in potential effectiveness degradations. However, in this work, we focus on \nt{the} safe setting, and leave unsafe settings to future work.}

\subsection{Relation to Dynamic Pruning Literature}\label{ssec:related}
\looseness -1 We now highlight parallels and contrasts with previous work in dynamic pruning. Firstly, we draw parallels in terms of nomenclature: items are documents; and sub-item ids are like terms, except that each item has a fixed number $M$ of sub-item ids, one from each split. These observations help us to position \jpqprune{} within the dynamic pruning literature. Indeed, existing dynamic pruning techniques cannot address this task, as our sequence embedding (query) can match with any sub-item id (term). 

\pageenlarge{2} Principle P1 suggests scoring items associated with highly-scored sub-item ids. This is similar to how optimised versions of the TAAT pruning score query terms in the decreasing order of Inverted Document Frequency~\cite[Sect. 3.2]{fntir2018}, allowing to find documents that are likely to be highly scored earlier. The use of a threshold $\theta$ from the top-ranked items is commonly deployed in DAAT dynamic pruning approaches (MaxScore, WAND) for the purposes of early terminating the scoring of documents. 

\looseness -1 Some other dynamic pruning approaches used impact-ordered postings lists~\cite[Ch. 5]{fntir2018}; on the surface this has some similarities to our work, however, the inclusion of splits, sub-item ids, etc. makes comparisons challenging. However, we note that Jia et al.~\cite{jia2010efficient} used updating upper bounds and a scoring terminating condition that bears resemblance to Principle P2. 
\cm{Principle P3 is novel, as we are not aware of dynamic pruning document retrieval literature considering batching of processing and the benefits of vectorisation.}
Moreover, we argue that \jpqprune{} is neither exclusively a DAAT nor a TAAT algorithm: like TAAT, it identifies a good sub-item id (term) to score next. Once that high-scoring sub-item id is identified, all items associated with that sub-item id are fully scored in DAAT fashion (as per Algorithm~1). In short, \jpqprune{} is a novel application of dynamic pruning ideas to RecJPQ scoring.

\section{Experimental Setup}\label{sec:exp}
Our experiments aim to answer the following Research Questions: 

\begin{enumerate}[font={\bfseries}, label={{\bfseries RQ\arabic*}}, leftmargin=*]
\item What is the effect of applying our \jpqprune{} algorithm on scoring \cm{efficiency}? \label{rq:efficiency} 
\item \sg{What is the effect of ranking cutoff $K$ on the efficiency of our \jpqprune{} algorithm}?
\label{rq:cutoff}
\item  \sg{What is the effect of varying the batch size on the efficiency our \jpqprune{} algorithm? \label{rq:batch_size}}
\end{enumerate}

\looseness -1 \nt{In the following, we detail the datasets used in our experiments (Section~\ref{ssec:datasets}), the used recommender models (Section~\ref{ssec:models}), and the measures we adopt to answer our research questions (Section~\ref{ssec:measures}).}

\begin{table}[tb]
    \centering
    \caption{Salient characteristics of experimental datasets.} \label{tb:datasets}\vspace{-1em}
    \small
    \sg{
    \begin{tabular}{lrrrrr}
    \toprule
    Dataset &  Users &  Items &  Interactions &  {Avg. length} \\
    \addlinespace[-0.1ex]
    \midrule
    Gowalla  &      86,168 &    1,271,638 &           6,397,903 &            74.24 \\
    Tmall &   473,376 &      2,194,464 &           34,850,828 &            73.62 \\
    \bottomrule
    \end{tabular}
}
\vspace{-1em}
\end{table}

\subsection{Datasets}\label{ssec:datasets}
\pageenlarge{2} \sgw{The focus of this paper is on large catalogues. Hence, in our experiments, we use datasets with some of the largest \craig{catalogues} available for academic research.} %
\looseness -1 \sgw{In particular}, we perform experiments using two large-scale sequence recommendation datasets, namely (i) Gowalla~\cite{choFriendship2011}, a point-of-interest check-in dataset, and (ii) Tmall~\cite{tianchiIJCAI16BrickandMortarStore2018}, \sgw{an e-commerce clicks dataset}.
Table~\ref{tb:datasets} provides the statistics of the datasets. Of note, both have larger numbers of items, i.e., 1.2M items in Gowalla, 2.1M items in Tmall,\footnote{\crc{We note that these datasets are among those with the largest number of items available for academic research. However, simulated experiments with  PQTopK have show that it remains efficient with larger-scale datasets with hunderds of million of items~\cite{petrovEfficientInferenceSubItem2024}. As PQTopK is the base algorithm for \jpqprune{}, we believe that findings from our experiments will also generalise to larger datasets.}} than conventional recommendation datasets such as MovieLens-1M, which has only 3K items. \crc{Indeed, for less than 30K items, even Transformer Default is very efficient, and further efficiency optimisation is not required as shown in~\cite{petrovEfficientInferenceSubItem2024}}. %
For both datasets, we remove sequences with less than 5 interactions and use the \sgw{\emph{temporal} "leave-one-out" (LOO) strategy for train/test split: we globally split the interactions by the timestamp so that the training period contains 90\% interactions; for test, we select the first interaction after the "Train/Test" splitting point.}
\sgw{Overall, our setting follows that from the RecJPQ~\cite{petrovRecJPQTrainingLargeCatalogue2024} and PQTopK publications~\cite{petrovEfficientInferenceSubItem2024} with the exception of using temporal LOO instead regular LOO.\footnote{\sgw{Regular LOO has recently been critiqued for potential data leakages~\cite{hidasiWidespreadFlawsOffline2023}.}}}

\subsection{Models \& Baselines}\label{ssec:models}%

\looseness -1 \cm{Given the large scale of our datasets, training \sgf{conventional {Trans-former}-based recommender} models would be challenging, particularly for BERT4Rec~\cite{BERT4Rec}, as it does not use negative sampling and instead computes a SoftMax over \textit{all} items in the catalogue for each position in the sequence. Instead, we use a RecJPQ variant of the popular SASRec model~\cite{SASRec}, which does use negative sampling,  denoted as SASRecJPQ. Further, to address the challenges caused by negative sampling, which is needed to train on large catalogues, we apply the gBCE \nt{loss} function~\cite{petrovGSASRecReducingOverconfidence2023} to both BERT4Rec and SASRec, denoting the final models gSASRecJPQ and gBERT4RecJPQ.}
\sasha{We align model training with the \sgt{RecJPQ paper}~\cite{petrovRecJPQTrainingLargeCatalogue2024}: we use a maximum sequence length of 200, \nt{and} 512-dimensional embeddings. For 1024 users, we select the last item before the train/test splitting timestamp as a validation set, which is used for early stopping model training if NDCG@10 did not improve for 200 epochs. \sgw{As recommended in~\cite{petrovRecJPQTrainingLargeCatalogue2024}, for RecJPQ we use $M=8$ splits and $B=256$ sub-ids (embeddings) per split.}} \cm{While our focus is on scoring efficiency, Appendix A  provides an overview of the effectiveness of the used models, and the time that the Transformer model takes to compute the sequence embedding.}

\looseness -1 For each model, we apply three methods for computing item scores: Transformer Default, which uses matrix multiplication, i.e. \sgw{the scoring procedure used in Transformer models by default\footnote{\sgw{We use a RecJPQ-based version of the models, so to use the default Transformer scoring, we obtain full item embeddings first using concatenation, i.e., as per Equation~\eqref{eq:embedding_reconstruct}. However, to ensure a fair comparison with Transformer Default, we do not include time spent on the reconstruction of the item embeddings in the scoring time.}}} (Equation~\eqref{eq:scores});  PQTopK~\cite{petrovEfficientInferenceSubItem2024}; and our \jpqprune{} method.

\looseness -1 \cm{We do not consider ANN implementations such as \sgw{FAISS}~\cite{FAISS}, as \sgf{they are not safe and cause} significantly reduced retrieval effectiveness.\footnote{Indeed, in preliminary experiments, we found that \sgw{FAISS} could result in 68\% degradations in NDCG@10 compared to SASRecJPQ.}} \sgw{Note that we only use RecJPQ-based models for our experiments, as training plain Transformer models without embedding compression is not feasible using consumer-grade hardware with catalogues of this size. For example, on the Tmall dataset, a full embedding table would require 2.2M items $\times$ 512 parameters per embedding $\times$ 4 bytes = 4.5GB GPU memory. \sgt{Considering the memory required for model gradients, moments, model parameters, and intermediate variables, we would need more than 24GB of GPU memory—exceeding what is currently available to us.}} This also prevents use of other PQ-based methods (such as those in FAISS), which require the training of full embeddings before compression.

\pageenlarge{2} \subsection{Measures}\label{ssec:measures}
\looseness -1 We are primarily focused on efficiency, which we analyse using model scoring time.\footnote{Note that any reported time is specific to our hardware: an AMD Ryzen 5950x CPU, 128Gb memory, no GPU acceleration.} \cm{Following~\cite{petrovEfficientInferenceSubItem2024}, our target environment considers only CPUs, i.e no GPU acceleration, at the inference time. Indeed, deploying a trained model on CPU-only hardware is often a reasonable choice for many high demand environments, considering the high costs associated with GPU accelerators.}
We exclude the time to obtain the sequence embedding through the Transformer layers, as this is a constant for all approaches \cm{(see Appendix~A}). Furthermore, we report \emph{median} (denoted \crc{mST}) scoring time instead of \emph{mean} because we observe that in our Tensorflow-based implementation, JIT compilation requires several iterations to warm up. Following the dynamic pruning literature~\cite{6228205,10.1145/3159652.3159676,jeon:2014,tonellotto:2013}, we also report 95th percentile scoring times, as dynamic pruning techniques can vary in the amount of pruning possible for different requests. Finally, we also report the number  of items scored by each algorithm, because the primary goal of \jpqprune{} is to avoid exhaustive scoring.

\section{Results} \label{sec:results}
We now address each of the research questions RQ1-RQ3 in turn.

\subsection{RQ1 - Overall Efficiency}
\looseness -1 We first analyse the efficiency of \jpqprune{} compared to the two baseline scoring methods, across three models (SASRecJPQ, gBERT4RecJPQ, gSASRecJPQ). Table~\ref{tb:results_table_new} reports the median and 95th percentile scoring times (in milliseconds) of \jpqprune{} method compared to baseline methods on the two experimental datasets, Gowalla and Tmall. For RQ1, we apply $K=10$ and a $BS = 8$, but \nt{we} investigate the impact of these parameters in RQ2 \& RQ3.

\looseness -1 From the table, it can be clearly seen that \nt{the Transformer Default baseline} involves excessive operations, resulting in \nt{large} median scoring times: \nt{more than 100ms on Gowalla and more than 200ms on Tmall.} Applying the existing method PQTopK, \sgw{which re-uses pre-computed sub-item id scores}, reduces \sgw{the median time} to around \nt{$9-16$ms -- an average speedup of $10.7\times$$-12.9\times$}. Both of these baselines apply no pruning, so 95\%tl times are very similar to the median.

On the other hand, applying our proposed \jpqprune{} dynamic pruning method, the median scoring times are reduced to 3-6ms, \nt{with a resulting speedup of 1.5$\times$-2.9$\times$ on Gowalla and 3.2$\times$-5.3$\times$ on Tmall compared to PQTopK}. This demonstrates the benefit of pruning at this scale, and focussing on splits that are more likely to result in the highest scored items being retrieved.

\pageenlarge{2} Furthermore, considering the  95\%tl times, we see that the pruning method can experience users that are difficult to prune; indeed, the 95\%tl scoring time for SASRecJPQ and gSASRecJPQ are \nt{1.2$\times$-2.4$\times$} slower than the median (although always faster than PQTopK). We examine pruning difficulty more in Section~\ref{sec:difficulty}. Similarly, the 95\%tl scoring time for gBERT4RecJPQ on Gowalla is slower than for PQTopK; as we will see in the next section, this model/dataset combination is more difficult for pruning. 

Overall, for RQ1, we find that \jpqprune{} can achieve improved median and 95\%tl scoring times, a reduction of up to \nt{$5.3\times$} compared to the median scoring time of the recent PQTopK approach \cm{(SASRecJPQ on Tmall: 16.72ms $\to$ 3.18ms), and up to $64\times$ compared to the Transformer Default baseline (204ms $\to$ 3.18ms).}

\begin{table}[tb]
\sg{
    \caption{\nt{Median (\crc{mST}) and tail (95\%tl)} \sgw{scoring} times of different scoring methods, \nt{in ms}. \sgw{The results are reported for ranking cutoff $K=10$. For \jpqprune{}, batch size $BS=8$.}}  \label{tb:results_table_new}\vspace{-.75em}
    \resizebox{\linewidth}{!}{
        \begin{tabular}{l rr rr rr}
\toprule
\multirow{2}{*}{Scoring Method} & \multicolumn{2}{c}{SASRecJPQ} & \multicolumn{2}{c}{gBERT4RecJPQ} & \multicolumn{2}{c}{gSASRecJPQ} 
\\

\cmidrule(lr){2-3}\cmidrule(lr){4-5}\cmidrule(lr){6-7}
& \crc{mST} & 95\%tl & \crc{mST} & 95\%tl &  \crc{mST} & 95\%tl   \\
\midrule
\addlinespace[-0.1ex]
\multicolumn{7}{c}{Gowalla} \\
\addlinespace[-0.5ex]
\midrule
Transformer Default & 123.61 & 126.77 & 123.24 &  126.75 & 123.87  & 126.88\\
 PQTopK & 10.19& 10.88 & 9.57 & \textbf{10.61} & 10.11  & 10.81\\
 RecPQPRune (ours) & \textbf{3.50}  & \textbf{8.51} & \textbf{6.42}& 19.79  & \textbf{4.59} & \textbf{7.99}\\
\midrule
\addlinespace[-0.1ex]
\multicolumn{7}{c}{Tmall} \\
\addlinespace[-0.5ex]
\midrule
Transformer Default & 204.18 & 210.48 & 205.67 & 210.76 & 206.38 & 210.89 \\
PQTopK & 16.72 & 18.26 & 16.66  & 17.76& 16.75  & 19.68\\
RecPQPRune (ours) & \textbf{3.18} & \textbf{4.59} & \textbf{5.11} & \textbf{6.53} & \textbf{5.20}  & \textbf{6.39}\\
\bottomrule
\end{tabular}

    }
}\vspace{-1.25em}
\end{table}

\subsection{RQ2 - Ranking Cutoff}

\looseness -1 The efficiency of existing document retrieval dynamic pruning methods are sensitive to the rank cutoff, $K$, because the threshold $\theta$ is obtained from the score of the current top $K$th ranked document, so retrieving fewer documents implies a higher threshold, causing less documents to be scored. Similarly, we expect reducing $K$ to also increase the efficiency of \jpqprune{}, for the same reasons. 

\begin{figure}
    \def\mstehiegh{0.12}

    \subfloat[Gowalla]{
        \includegraphics[trim=0.5cm 0.3cm 0cm 0.5cm,width=0.24\textwidth]{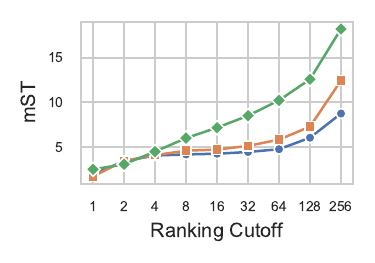}
    }
    \subfloat[Tmall]{   
        \includegraphics[trim=0.5cm 0.3cm 0cm 0.5cm,width=0.24\textwidth]{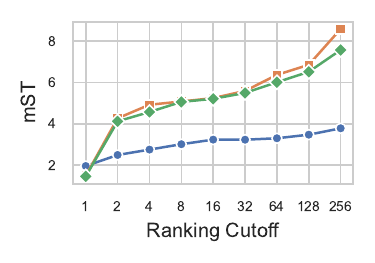}
    }

    \vspace{-2mm}\subfloat{
        \includegraphics[width=.7\linewidth]{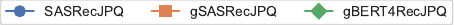}
    }\vspace{-2mm}
    \caption{\crc{Ranking cutoff vs. median scoring time (mST, ms).}}\label{fig:cutoff}\vspace{-2em}
\end{figure}

\looseness -1 To analyse this, we turn to Figure~\ref{fig:cutoff}, which plots the scoring time of the various models on both Gowalla and Tmall datasets as the rank cutoff $K$ is varied between 1 and 256. Its clear from the figures that, as expected, scoring time indeed increases as the rank cutoff increases. There appears to be a marked increase in scoring time between 128 and 256 retrieved items; however this is more an artifact of the logarithmic scale in the x-axis of the figures. We also observe some variance between the different models: SASRec is consistently the fastest model on both datasets as $K$ is varied; gBERT4RecJPQ and gSASRecJPQ are typically slower (with gBERT4RecJPQ being slower than gSASRecJPQ for Gowalla). We examine the relative difficulty of pruning different models later in Section~\ref{sec:difficulty}. 

To summarise for RQ2, we find that, as expected, decreasing the rank cutoff decreases the scoring time. While this is expected from the existing dynamic pruning literature, it is a characteristic not previously observed Transformer-based recommender systems, where the Transformer Default method requires all items to be scored and then sorted for a given sequence embedding. 

\subsection{RQ3 - Batch Size} \label{ssec:exp:bs}
\begin{figure}[tb]
\def\batchsizescale{0.23}
    \subfloat[Gowalla, mST]{
        \includegraphics[trim=0.5cm 0.2cm 0cm 0.5cm, width=\batchsizescale\textwidth,]{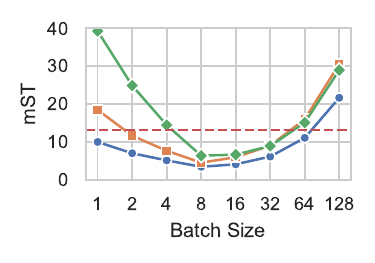}
    }%
    \subfloat[Tmall, mST]{
        \includegraphics[trim=0.5cm 0.2cm 0cm 0.5cm, width=\batchsizescale\textwidth]{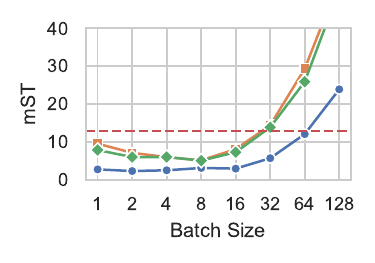}
    }
    
    \subfloat[Gowalla, \% scored items]{
       \includegraphics[trim=0.5cm 0.2cm 0cm 0.5cm, width=\batchsizescale\textwidth]{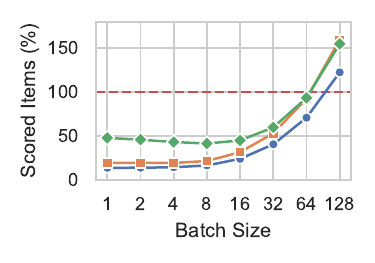}
    }
    \subfloat[Tmall, \% scored items]{
       \includegraphics[trim=0.5cm 0.2cm 0cm 0.5cm, width=\batchsizescale\textwidth]{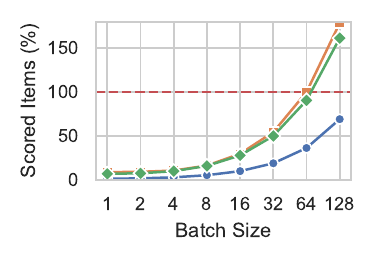}
    }
    \vspace{-.5\baselineskip}
    \subfloat{
        \includegraphics[width=.9\linewidth]{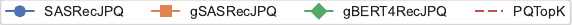}
    }
    \caption{Effect of the batch size on median scoring time \sgw{(mST, top)} and the number of scored items (bottom) on Gowalla and Tmall in \jpqprune{} for ranking cutoff $K$=10. PQTopK baseline is averaged across the three models.}\label{fig:bs:mrt}\vspace{-2em}
\end{figure}

Finally, we consider the other parameter of \jpqprune{}, namely the batch size, $BS$, that controls how many sub-item ids are processed concurrently. Figure~\ref{fig:bs:mrt} reports the median scoring time as batch size is varied, on both Gowlla and Tmall datasets \sgw{as well as the percentage of processed items during scoring}. We present one line for each RecJPQ-based model, while for PQTopK we report an average across the three models (which are very close - see Table~\ref{tb:results_table_new}).

\pageenlarge{2} \looseness -1 From Figure~\ref{fig:bs:mrt} (a) \& (b), it is clear that for median scoring time there is a sweet spot for batch size $BS$ - around 8 on both datasets and all models. \sgw{The existence of this sweet spot confirms our Principle P3 and demonstrates the importance of batched processing.} Smaller batch sizes typically result in increased scoring times - particularly so for gBERT4RecJPQ on Gowalla. The increases in scoring time suggest higher overheads from smaller batch sizes e.g.\ more applications of PQTopK and merge operations (lines \ref{alg:jpqrune1:score} \& \ref{alg:jpqprune1:update_topk} in Algorithm~1). High batch sizes also exhibit increased scoring times, as \sgw{more items are scored than needed to achieve the pruning condition}.

To quantify this behaviour, Figures~\ref{fig:bs:mrt} (c) \& (d) report the percentage of items scored for different models as batch size is varied. From the figure, it can be seen that the percentage of items scored is typically reduced as batch size reduces - this makes sense, as fewer sub-items are selected at each main loop iteration. Indeed, it is expected that scoring time is heavily correlated with scored items~\cite{10.1145/2348283.2348367,10.1145/3389795} - and therefore, the increased scoring times for small batch sizes come from the overheads, as discussed above. The exception here is again gBERT4RecJPQ on Gowalla. This outlier model is discussed further in Section~\ref{sec:difficulty}. \sgw{An interesting observation from the figures is that the percentage of items scored can exceed 100\%. Indeed, \jpqprune{} does not \sgt{maintain a} set of already \sgt{scored} items, and the same item may be scored repeatedly when the algorithm processes different sub-ids associated with this item. While it could be possible to \sgt{maintain a} set of already processed items and process every item only once, our initial experiments showed that the overhead associated with maintaining such a set and checking every item is larger than the cost of repeated scoring of the items.}

\pageenlarge{2} \sgw{Overall, in answer to RQ3, we find that setting the batch size appropriately is important to achieve efficiency gains compared to the baselines. As suggested by Principle P3, setting the batch size too small (e.g. 1) increases computational overhead, while setting a high batch size increases the number of scored items. In our experiments, we find the "sweet spot" for batch size is at value 8, which we recommend as a default value for \jpqprune{}.}

\section{Pruning Difficulty}\label{sec:difficulty}

\begin{figure}
    \subfloat[Fast (1ms, 2.9\% items scored, 1 iteration), gBERT4RecJPQ]{
        \includegraphics[trim=0cm 1cm 0cm 0.5cm,clip=true,width=0.95\linewidth]{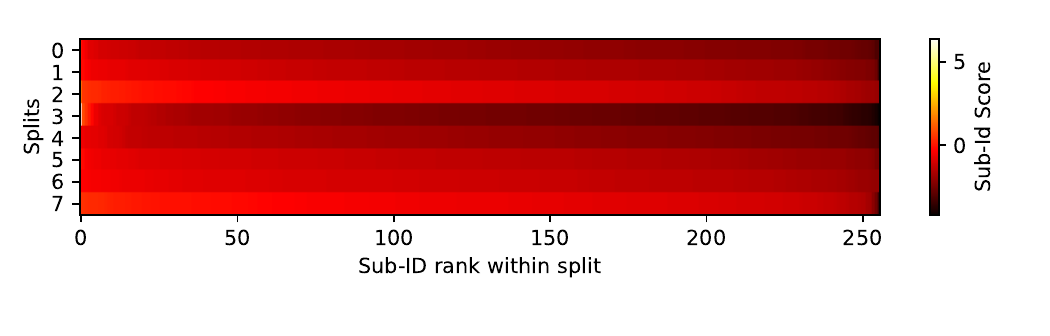}
    }\vspace{-0.7em}
    
    \subfloat[Average (6ms, 32\% items scored, 11 iterations), gBERT4RecJPQ]{
        \includegraphics[trim=0cm 1cm 0cm 0.5cm,clip=true,width=0.95\linewidth]{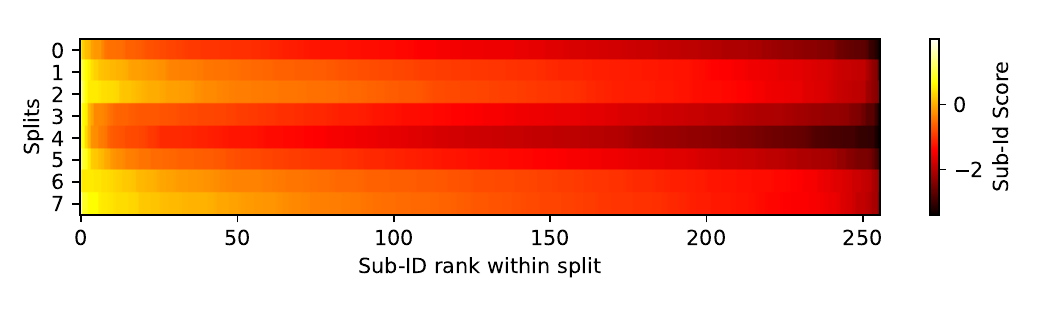}
    }\vspace{-0.7em}

    \subfloat[Average (4ms, 23\% items scored, 8 iterations), SASRecJPQ]{
        \includegraphics[trim=0cm 1cm 0cm 0.5cm,clip=true,width=0.95\linewidth]{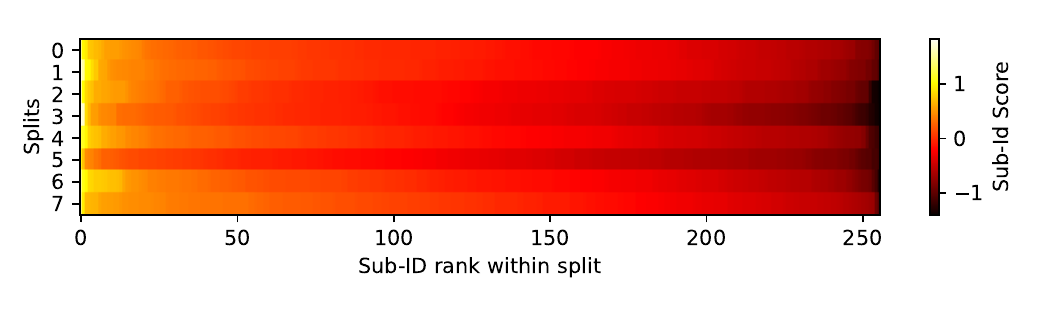}
    }\vspace{-0.7em}
    
    \subfloat[Slow (91ms, 102\% items scored, 35 iterations), gBERT4RecJPQ]{
        \includegraphics[trim=0cm 1cm 0cm 0.5cm,clip=true,width=0.95\linewidth]{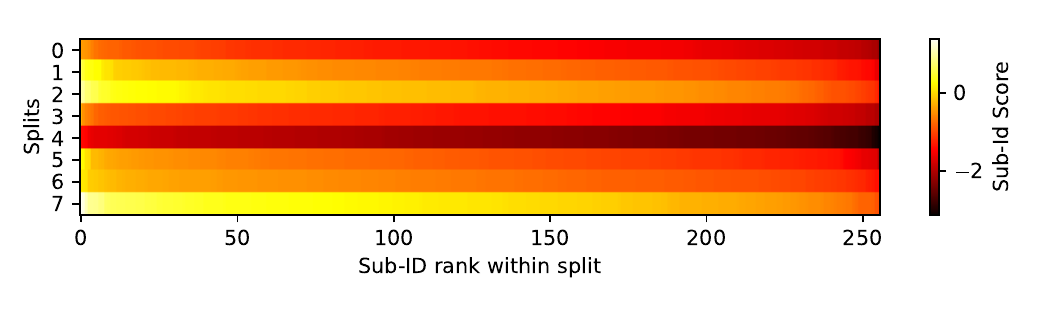}
    }\vspace{-0.7em}
    \caption{Sub-item id scores for three users: fast (a, gBERT4RecJPQ), average (b, gBERT4RecJPQ) \& (c, SASRecJPQ), and slow (d, gBERT4RecJPQ).
    Y-axis represent splits, x-axis represent rank of the sub-item id within split (ordered by score) and colour represents the score of the sub-item id.} \label{fig:heatmaps}
\end{figure}

\looseness -1 \sg{In our experiments \jpqprune{} shows improvements over both Transformer Default and PQTopK. \sgf{However, from Table~\ref{tb:results_table_new}, we can also see that the margin between the \crc{median} and \crc{the 95th percentile} of scoring time is larger for \jpqprune{} compared to baselines.  Indeed, for the baselines, the scoring complexity does not depend on the user. In contrast, for \jpqprune{}, the number of iterations depends on the user -- for most users, the scoring time is low, \sgw{but }there are some users for which the algorithm requires many iterations.}}

 \sg{To better understand what makes users more or less suitable for \jpqprune{}, we visualise sub-item id scores using on the Gowalla dataset in Figure~\ref{fig:heatmaps} for three different users for gBERT4RecJPQ: a \textit{fast/average/slow} user, for which the algorithm needs /1/6/91 ms.
 \sgw{To illustrate the difference between the models, we also include a visualisation of the same \emph{average} user with the SASRecJPQ model, where scoring requires 4ms.} The captions also report the percentage of items scored \sgw{for these 3 users, which varies considerably, between 2.9\% and 102\% (recall that some items may be scored repeatedly, so the percentage of scored items can exceed 100\%)} implying that the {\em pruning difficulty}~\cite{10.1145/2348283.2348367} of these users varies considerably.  Interestingly, pruning difficulty can be inferred from inspection of Figure~\ref{fig:heatmaps}, as there is a clear difference between the sub-item id scores distribution for these three users. Indeed, for the fast user, there are very few sub-item ids with high scores, and all are located in the same 3\textsuperscript{rd} split, so the algorithm can identify most highly scored items very rapidly (\sgw{just a single iteration of the algorithm}).  For the average user, we see a more balanced distribution of sub-item id scores for \sgw{for both gBERT4RecJPQ and SASRecJPQ}, but there are still few high-scored sub-item ids. For the slow user, most of the sub-item ids in the 7\textsuperscript{th} and 2\textsuperscript{nd} splits have high scores, forcing the algorithm to process most of the sub-item ids from these splits.} \sgw{Finally, considering the difference between the visualisations for the average user between gBERT4RecJPQ and SASRecJPQ, we see that highly-scored sub-ids for the latter are more concentrated to in the left side, making gSASRec more efficient for this user.}
 
\pageenlarge{2} Overall, these differences highlight an important property of \jpqprune{}: \emph{it is most efficient when the model is confident, such that the sequence emebdding only has to \cm{score} a few highly scored sub-item ids \cm{before terminating}}. This insight opens a path for future research: the model can be \emph{trained} to make pruning more efficient. %

\begin{table}[tb]
\caption{Salient characteristics of the used experimental models: effectiveness, time to calculate the sequence embedding $\phi$ by the Transformer model, and model checkpoint size.}\label{tab:expmodels}\vspace{-1em}
\resizebox{\linewidth}{!}{
\begin{tabular}{ll ccc}
\toprule
Version & Model & NDCG@10 & \makecell{Transformer\\ to $\phi$ (ms)} & \makecell{Checkpoint\\Size (MB)} \\ 
\midrule
\addlinespace[-0.1ex]
\multicolumn{5}{c}{Gowalla -- 1.2M items} \\
\addlinespace[-0.5ex]
\midrule

\multirow{3}{*}{Ours} & SASRecJPQ    & 0.1142  & 24.81 & 91\\         

  & gBERT4RecJPQ & 0.1718  & 37.27   & 162 \\
& gSASRecJPQ   & 0.1667  & 24.55     & 90 \\
\midrule
\multirow{2}{*}{Reported~\cite{petrovRecJPQTrainingLargeCatalogue2024,petrovEfficientInferenceSubItem2024}}     & SASRecJPQ    & 0.1220   & 24.67                                                           & 69   \\
& SASRec       & 0.1100   & 24.67 & 3200  \\ 
\midrule
 \addlinespace[-0.1ex]
 \multicolumn{5}{c}{Tmall -- 2.2M items} \\
 \addlinespace[-0.5ex]
\midrule
\multirow{3}{*}{Ours} & SASRecJPQ  & 0.0049  & 24.67    & 185\\
 & gBERT4RecJPQ   & 0.0155  & 37.62       & 258 \\
 & gSASRecJPQ   & 0.0134  & 24.71      & 186 \\                      
\bottomrule
\end{tabular}}\vspace{-1em}
\end{table}

 \section{Conclusions}\label{sec:conc}
\looseness -1 We introduced \jpqprune{} as a novel dynamic pruning algorithm for scoring RecJPQ-based sub-item representations within embedded sequential recommender systems. \jpqprune{} is inspired by  dynamic pruning from document retrieval but tackles a completely new and more recent problem faced by Transformer-based recommendation models asked to predict over large item catalogues. Our \sgw{experiments} demonstrated the efficiency benefits of the approach: for instance, on the Tmall dataset with 2.2M items, we can reduce model scoring time by 64$\times$ compared to the Transformer Default baseline, and 5.3$\times$ compared the recent PQTopK approach. This is achieved while being safe-up-to-rank-$K$, i.e., no impact on effectiveness upto rank $K$; efficiency is further enhanced as $K$ is reduced. %
\crc{Our future work will consider unsafe configurations of  \jpqprune{}.
Furthermore,} we believe that PQ-based implementations of Transformer models combined with \jpqprune{} also have applications to generative retrieval models~\cite{pradeepHowDoesGenerative2023} and recommender systems~\cite{rajputRecommenderSystemsGenerative2023, GPTRec}, where the Transformer generates the ids to retrieve. We leave testing of our method for generative settings to future work.\pageenlarge{2}

\appendix
\renewcommand\thesection{APPENDIX \Alph{section}}

\section{Model Effectiveness etc.}\label{ax:ndcg}
    Table~\ref{tab:expmodels} provides an overview of the effectiveness and efficiency of our RecJPQ-based models. \sgw{Note that we use a full version of the Tmall (aka. Taobao) dataset with 2.2M items; this is a larger number of items than usually used for this dataset~\cite{maModelingPeriodicPattern2020,chenDynamicCoattentionNetwork2019, xuRecurrentConvolutionalNeural2019}, but means our reported effectiveness measures are not comparable the numbers reported in these papers}.
For Gowalla, we also include the results for plain SASRec, i.e. w/o RecJPQ, obtained from~\cite{petrovRecJPQTrainingLargeCatalogue2024,petrovEfficientInferenceSubItem2024}. On Gowalla, it can be seen that applying RecJPQ to SASRec results in much smaller models in terms of checkpoint size (3200MB vs.\ less than 169MB), as well as increased effectiveness, due to regularisation of the model. On both datasets, gBERT4RecJPQ and gSASRecJPQ improve NDCG@10 due to the use of the gBCE loss function. The time taken for the Transformer models to compute the sequence embedding $\phi$ does not considerably vary across datasets, although gBERT4RecJPQ (an Encoder-based model) is slower than the SASRec models, which are Decoder-based. \sgf{In our experiments, we verified that all scoring methods produce identical NDCG@10.}

\vspace{0.5\baselineskip}
\looseness -1 \noindent \textbf{\large Acknowledgments.}
Tonellotto acknowledges the Spoke ``FutureHPC \& BigData'' of the ICSC – Centro Nazionale di Ricerca in HPC, Big Data and Quantum Computing, the FoReLab project (Departments of Excellence), and the NEREO PRIN project of the Italian Ministry of Education and Research (Grant no. 2022AEFHAZ).

\FloatBarrier

\bibliographystyle{ACM-Reference-Format}
\balance
\bibliography{bib,doceng,references}
\end{document}